\documentclass[prl,twocolumn,groupedaddress]{revtex4-1}
\catcode`\@=11
\def\lsim{\mathrel{\mathpalette\@versim<}}
\def\gsim{\mathrel{\mathpalette\@versim>}}
\def\@versim#1#2{\vcenter{\offinterlineskip
\ialign{$\m@th#1\hfil##\hfil$\crcr#2\crcr\sim\crcr } }}
\catcode`\@=12

\catcode`@=12
\usepackage[export]{adjustbox}	
\usepackage{float}
\usepackage{caption}
\usepackage{subcaption}
\usepackage{epsfig,bbm}
\usepackage{amsmath}
\usepackage{slashed}
\usepackage{tikz}
\usepackage{bm}
\usepackage{amssymb}
\usepackage{mathtools}
\usepackage{graphicx}
\newcommand{\be}{\begin{equation}}
\newcommand{\ee}{\end{equation}}
\newcommand{\bea}{\begin{eqnarray}}
\newcommand{\eea}{\end{eqnarray}}

\begin{document}
\thispagestyle{empty}
\begin{flushright}
\end{flushright}

\begin{center}
\title{Small Neutrino Masses from a Decoupled Singlet Scalar Field
}
\author{Jongkuk Kim}
\email[]{jongkuk.kim927@gmail.com}
\affiliation{Department of Physics, Chung-Ang University, Seoul
06974, Korea}
\author{Seong-Sik Kim}
\email[]{sskim.working@gmail.com}
\affiliation{Department of Physics, Chung-Ang University, Seoul
06974, Korea}
\author{Hyun Min Lee}
\email[]{hminlee@cau.ac.kr}
\affiliation{Department of Physics, Chung-Ang University, Seoul
06974, Korea}
\author{Rojalin Padhan}
\email[]{rojalinpadhan2014@gmail.com}
\affiliation{Department of Physics, Chung-Ang University, Seoul
06974, Korea}

\begin{abstract}
We propose a unified solution with $Z_4$ discrete symmetry for small neutrino masses and stability of dark matter. The Standard Model is extended with an inert doublet scalar, a dark singlet scalar, a spurion scalar and right-handed neutrinos, which all transform nontrivially under $Z_4$. After the $Z_4$ symmetry is broken to $Z_2$ by the VEV of the spurion, much below the mass scale of the dark singlet scalar, a small lepton number violating coupling for the inert doublet is generated at tree level, so small neutrino masses are obtained at one-loops for relatively light new fields. We discuss the important roles of the $Z_4$ symmetry for neutrino masses, dark matter physics and thermal leptogenesis. 
 
\end{abstract}
\maketitle
\end{center}

\section{ Introduction}
Neutrino oscillations \cite{ParticleDataGroup:2022pth} indicate neutrinos masses of $m_\nu\sim 0.05\,{\rm eV}$. Moreover, the bounds from neutrinoless double beta decay experiments \cite{KATRIN:2021uub} and Cosmic Microwave Background anisotropies and Baryon Acoustic Oscillations \cite{Planck:2018vyg} are $m_{{\bar\nu}_e}<0.8\,{\rm eV}$  and \mbox{$\sum_i m_{\nu_i}<0.12\,{\rm eV}$} at 95\% CL, respectively. In order to account for small neutrino masses, it is possible to extend the Standard Model (SM) by the dimension-5 operators for neutrino masses with a cutoff scale beyond the electroweak scale or light sterile neutrinos coupled to neutrinos much below the electroweak scale \cite{Cai:2017jrq}.

Small neutrino masses can be obtained by the mixing between active neutrinos and heavy Majorana neutrinos \cite{Minkowski:1977sc,Yanagida:1979as,Gell-Mann:1979vob,Mohapatra:1979ia}.  
Alternatively, small neutrino masses can be generated by loop corrections, for instance, due to right-handed (RH) neutrinos and an extra doublet scalar \cite{Ma:2006km}, the latter of which serves as a dark matter candidate \cite{Barbieri:2006dq,LopezHonorez:2006gr}. However, there is still a need of very small parameters in obtaining small neutrino masses in similar models \cite{Cai:2017jrq}.

In this article, we propose a novel model with a $Z_4$ discrete symmetry for the common origin of small neutrino masses and stability of dark matter. In our model, there are one inert doublet scalar, three right-handed neutrinos, a real spurion scalar and a singlet complex scalar field, all transforming nontrivially under $Z_4$. The lepton number is respected in the renormalizable interactions due to $Z_4$. Then, only after the $Z_4$ symmetry is broken to $Z_2$ due to the VEV of the spurion field, the lepton number is broken, so neutrino masses are generated at one-loops. Moreover, the lightest neutral dark scalar in the model becomes a dark matter candidate. The spurion scalar can be regarded as a pseudo-Goldstone boson coming from the breaking of a global $U(1)$ symmetry into $Z_4$ at a higher scale, and $Z_4$ can be further down to $Z_2$ at a low scale.  

Taking the decoupling limit where the singlet scalar field $S$ becomes heavier than the $Z_4$ breaking scale, we show how a small mass splitting between the neutral components of the inert doublet arises, in turn giving rise to small neutrino masses at one-loops. In the present case, we argue that large masses of extra fields or small Yukawa couplings are not necessary, unlike the case with the $Z_2$ symmetry only.
We discuss the roles of the new singlet scalars and constraints for neutrino masses, dark matter and thermal leptogenesis.

\section{ Discrete symmetry and dark matter scalars}
We introduce a complex singlet scalar $S$ and one extra electroweak doublet scalar, $H_2$, other than the SM Higgs doublet $H_1$, and three right-handed (RH) neutrinos, $N_{R,i}$, with $i=1,2,3$, and a spurion real scalar field $\varphi$, beyond the SM. We impose a $Z_4$ symmetry for the extra fields such that $\phi_i \to e^{i Q_X\pi/2} \phi_i$ with $\phi_i=S, \varphi, H_1, H_2, N_R$ and $Q_X(\phi_i) = (+1,  +2, 0, +1,+1)$. But, the SM fermions are neutral under $Z_4$ so the model is different from the flavor-dependent $Z_4$ symmetry introduced in Ref.~\cite{Babu:2013pma}. The $Z_4$ symmetry can be embedded into a global $U(1)$ Peccei-Quinn symmetry which is anomalous in the presence of the couplings to the SM quarks or extra quarks, but we don't pursue this possibility in detail in the current work.  As the subgroup of the $Z_4$ symmetry, we can identify $Z_2$ parities as $Z_2(\phi_i)=(-,+,+,-,-)$. The spurion field $\varphi$ transforms by $\varphi\to -\varphi$ under the $Z_4$ symmetry but it is invariant under the $Z_2$ subgroup. 

Now we consider the relevant interaction Lagrangian for the model as
\bea
{\cal L}_{\rm int}&=& -\sqrt{2}\kappa S^\dagger H^\dagger_1 H_2 -\sqrt{2}\lambda'_{S\varphi}  \varphi S H^\dagger_1 H_2\nonumber  \\
&&  -y_{N,ij} {\bar l}_i {\widetilde H}_2 N_{R,j} -\frac{1}{2}\lambda_{N,ij}\varphi \overline{N^c_{R,i}} N_{R,j} +{\rm h.c.} \label{Lint}
\eea
with ${\widetilde H}_2=i\sigma_2H^*_2$.
It is remarkable that $\frac{1}{2}\lambda_5 (H^\dagger_1 H_2)^2$ and its hermitian conjugate, vanish due to the $Z_4$ symmetry, and there is no bare mass terms for right-handed neutrinos due to the lepton number conservation.
There are extra $Z_4$ invariant interaction terms in the potential such as $\mu \varphi S^2, \frac{1}{2}\lambda_{H_i\varphi}\varphi^2 |H_i|^2$($i=1,2$), etc, as shown in the appendix, but it is sufficient to focus on eq.~(\ref{Lint}) for our discussion below.

After $\varphi$ gets a VEV by $\langle\varphi\rangle=v_\varphi$, $Z_4$ is broken to $Z_2$ and we obtain the  $Z_2$ invariant effective interactions and the mass terms for right-handed neutrinos,
\bea
{\cal L}_{\rm int, eff}&=& -\sqrt{2}\kappa S^\dagger H^\dagger_1 H_2 -\sqrt{2}\kappa' S H^\dagger_1 H_2 \nonumber  \\
&&  -y_{N,ij} {\bar l}_i {\widetilde H}_2 N_{R,j} -\frac{1}{2}M_{N,i} \overline{N^c_{R,i}} N_{R,i} +{\rm h.c.} \label{Leff}
\eea
where $\kappa'\equiv\lambda'_{S\varphi}  v_\varphi$ is the $Z_4$ breaking effective interaction and $M_{N,i}\equiv \lambda_{N,ii} v_\varphi$ are the masses for RH neutrinos. Here, we assumed the basis where $\lambda_{N,ij}$ is diagonalized.
After electroweak symmetry breaking with $\langle H_1\rangle=(0,v_H)^T/\sqrt{2}\neq 0$, there appear mass mixings between dark matter scalars, namely, $S$ and the neutral components of $H_2$, so nonzero neutrino masses can be generated at one-loops, which will be discussed later. When the SM Higgs and the spurion field $\varphi$ are expanded around the VEVs as $ H_1=(0,v_H+h)^T/\sqrt{2}$ and $\varphi=v_\varphi+\rho$, the dark Higgs $\rho$ can mix with the SM Higgs $h$, so the mass eigenstates become $h_1=c_\alpha \, h+s_\alpha\, \rho$ and $h_2=-s_\alpha \, h+c_\alpha\, \rho$, with $c_\alpha\equiv \cos\alpha, s_\alpha\equiv \sin\alpha$.

We remark that $S$ and $H_2$ do not get VEVs, leaving a $Z_2$ symmetry unbroken.
$S$, $H_2$ and $N_{R,i}$ are odd under the remaining $Z_2$, which ensures the stability of the lightest neutral $Z_2$-odd particle among them as a dark matter candidate. For small neutrino masses without introducing small parameters, we consider the case where the singlet scalar $S$ is decoupled, so either the lighter neutral component of $H_2$ or the lightest RH neutrino can be a dark matter candidate.

\section{ Decoupled scalar and inert doublet scalar masses}
\begin{figure}[t]
\centering
\includegraphics[width=0.18\textwidth,clip]{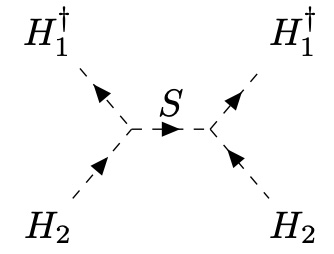}\qquad\qquad
\includegraphics[width=0.15\textwidth,clip]{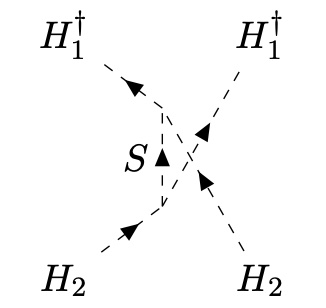} 
\caption{Feynman diagrams contributing to $\lambda_{5,{\rm eff}}$.
}
\label{fig:tree}
\end{figure}

From the effective interactions in eq.~(\ref{Leff}), we can take the limit of integrating out the singlet scalar $S$ whose squared mass becomes $m^2_S\gg \kappa \kappa'$. As a result, we obtain a renormalizable but $Z_4$ breaking effective interaction for two Higgs doublets from $\kappa, \kappa'$ terms, at tree level as shown in Fig.~\ref{fig:tree}, in the following form,
\bea
{\cal L}_{\rm eff}=-\frac{1}{2}\lambda_{5,{\rm eff}} (H^\dagger_1 H_2)^2+{\rm h.c.} \label{lamb5}
\eea
with
\bea
\lambda_{5,{\rm eff}}=-\frac{4\kappa\kappa'}{m^2_S}. \label{lamb52}
\eea
Here, $\kappa$ is a $Z_4$ invariant dimensionful coupling, so it can be of order $m_S$. But, $\kappa'$ is a $Z_4$ breaking coupling due to the VEV of the spurion field $\varphi$, so it is naturally suppressed as compared to $m_S$, that is, $\kappa'=\lambda'_{S\varphi} v_\varphi\ll m_S$. 
Therefore, the effective $\lambda_{5,{\rm eff}}$ term in eq.~(\ref{lamb52}) is also suppressed as a consequence of the decoupled scalar.

For a small $\lambda_5$ term in the effective theory below the threshold of the singlet scalar $S$, we obtain the masses for the components of the inert doublet, $H_2=(H^+,\frac{1}{\sqrt{2}}(H_0+iA_0))^T$, as follows,
\begin{align}
m^2_{H^\pm} &= m^2_2 +\frac{1}{2 }\lambda_3 v^2_H, \\
m^2_{H_0} &=   m^2_2 + \frac{1}{2} \left( \lambda_L+\lambda_{5,{\rm eff}} \right)v^2_H  , \\
m^2_{A_0}&= m^2_2 + \frac{1}{2} \left( \lambda_L-\lambda_{5,{\rm eff}} \right)v^2_H 
\end{align}
where $\lambda_L\equiv \lambda_3+\lambda_4$, with $\lambda_3,\lambda_4$ being mixing quartic couplings in the scalar potential, $\lambda_3|H_1|^2|H_2|^2+\lambda_4(H^\dagger_1 H_2)(H^\dagger_2 H_1)$, and $m^2_2$ is the squared mass parameter for the inert doublet $H_2$.
Henceforth, we choose $\lambda_{5,{\rm eff}} >0$ without loss of generality such that $A_0$ is a dark matter candidate. 
The most general results for the mass eigenvalues and the mixing angles with the non-decoupled singlet scalar $S$ are shown in eqs.~(\ref{HCmass}), (\ref{Hmasses}) and (\ref{Amasses}), in the appendix.

In the decoupling limit of the singlet scalar $S$, the small effective $\lambda_5$ term gives rise to a mass splitting between two real neutral scalars of $H_2$ is suppressed as
$
\Delta \equiv m_{H_0}-m_{A_0}\simeq \frac{\lambda_{5,{\rm eff}} v^2_H}{2m_0}, 
$
with $m^2_0\equiv (m^2_{H_0} +m^2_{A_0} )/2$. 
So, we can avoid the strong bound on DM-nucleon scattering with $Z$ boson exchange, as far as $|\Delta|\gtrsim 100\,{\rm keV}$, which sets a lower bound on $\lambda_{5,{\rm eff}}$ as $\lambda_{5,{\rm eff}}\gtrsim 3.3 \times 10^{-7} (m_0/100\,{\rm GeV})(|\Delta|/100\,{\rm keV})$.
Then, for $\kappa\sim m_S$, we need $\kappa'/m_S\gtrsim 8\times 10^{-8}$ for $m_0\sim 100\,{\rm GeV}$ and $|\Delta|\gtrsim 100\,{\rm keV}$.

\begin{figure}[t]
\centering
\includegraphics[width=0.30\textwidth,clip]{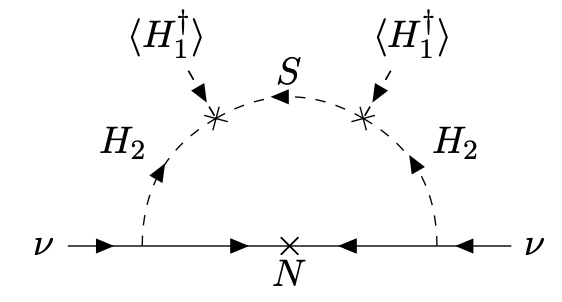}
\caption{Feynman diagrams contributing to neutrino masses at one-loop.
}
\label{fig:loop}
\end{figure}

\vspace{0.3cm}

\section{ Decoupled scalar and small neutrino masses}
In the decoupling limit of  the singlet scalar $S$, from the combination of the Yukawa interactions for right-handed neutrinos and a small effective $\lambda_{5,{\rm eff}}$ term in eq.~(\ref{lamb5}), we obtain the one-loop corrections for neutrino masses from the Feynman diagram shown 
in Fig.~\ref{fig:loop} as
\bea
({\cal M}_\nu)_{ij} &\simeq& \frac{\lambda_{5,{\rm eff}}v^2_H}{16\pi^2}\sum_k \frac{y_{N,ik}y_{N,jk}M_{N,k}}{m^2_0-M^2_{N,k}}\nonumber\\
&&\times\bigg[1-\frac{M^2_{N,k}}{m^2_0-M^2_{N,k}}\ln\frac{m^2_0}{M^2_{N,k}}\bigg]. \label{numass}
\eea
The above results match with those due to a nonzero small $\lambda_5$ in the case without a $Z_4$ symmetry in Ref.~\cite{Ma:2006km}, and the same results can be obtained from the decoupling limit of the general results in eq.~(\ref{genumass}) in the appendix. 
Then, when  the RH neutrinos are heavier than the inert doublet Higgs, namely, $M_{N,k} \gtrsim m_0$, the above expression with eq.~(\ref{lamb52}) gets simplified to
\bea
({\cal M}_\nu)_{ij} \simeq \frac{\lambda_{5,{\rm eff}} v^2_H}{16\pi^2} \sum_k \frac{y_{N,ik}y_{N,jk}}{M_{N,k}}\bigg[\ln \frac{M^2_{N,k}}{m^2_0}-1\bigg]. 
\eea
Therefore, for neutrino masses, $m_\nu\lesssim 0.1\,{\rm eV}$, choosing $y_{N,ij}={\cal O}(1)$ and $M_{N,k}\sim 10^9\,{\rm GeV}$, we can set the upper bound on $\lambda_{5,{\rm eff}}$ as $\lambda_{5,{\rm eff}}\lesssim 8\times 10^{-6}$, which is consistent with the lower bound on $\Delta$ from the direct detection. In other words, we need $\kappa'/m_S\lesssim 10^{-6}$ for $\kappa\sim m_S$ and $M_{N,k}\lesssim \kappa'\ll m_S$ for  $\lambda_{N,kk}\lesssim \lambda'_{S\varphi}$. Then, we can keep the RH right-handed neutrinos in the effective theory while the singlet scalar $S$ is decoupled. So, for a small $\lambda_{5,{\rm eff}}$, it is sufficient to take $m_S\gtrsim M_{N,k}\sim10^9\,{\rm GeV}$, unless we tune the Yukawa couplings $y_{N,ij}$ to small values.

In the case with $m_0\gtrsim M_{N,k}$, the lightest RH neutrino is a dark matter candidate \cite{Ibarra:2016dlb} whereas the scalars of the inert doublet decay into a pair of the SM lepton and the RH neutrino through the Yukawa interactions. In this limit, eq.~(\ref{numass}) becomes approximated to
$
({\cal M}_\nu)_{ij} \simeq \frac{\lambda_{5,{\rm eff}}v^2_H}{16\pi^2m^2_0}\sum_k y_{N,ik}y_{N,jk}M_{N,k}. 
$
For instance,  for $m_\nu\lesssim 0.1\,{\rm eV}$, $y_{N,ij}={\cal O}(1)$, $m_0\sim 100\,{\rm GeV}$ and $M_{N,k}\sim 10\,{\rm MeV}$, we can set the upper bound, $\lambda_{5,{\rm eff}}\lesssim 8\times 10^{-10}$ or $|\Delta|\lesssim 0.2\,{\rm keV}$. But, there is no more lower bound on $\lambda_{5,{\rm eff}}$ because the scalars in the inert doublet decay. A small effective $\lambda_{5,{\rm eff}}$ is achievable if $m_S\gtrsim 10^9\,{\rm GeV} (\lambda'_{S\varphi}/\lambda_{N,kk})$ for $v_\varphi= 10\,{\rm MeV}/\lambda_{N,kk}$ for $\kappa\sim m_S$.  

\begin{figure}[t]
\centering
\includegraphics[width=0.45\textwidth,clip]{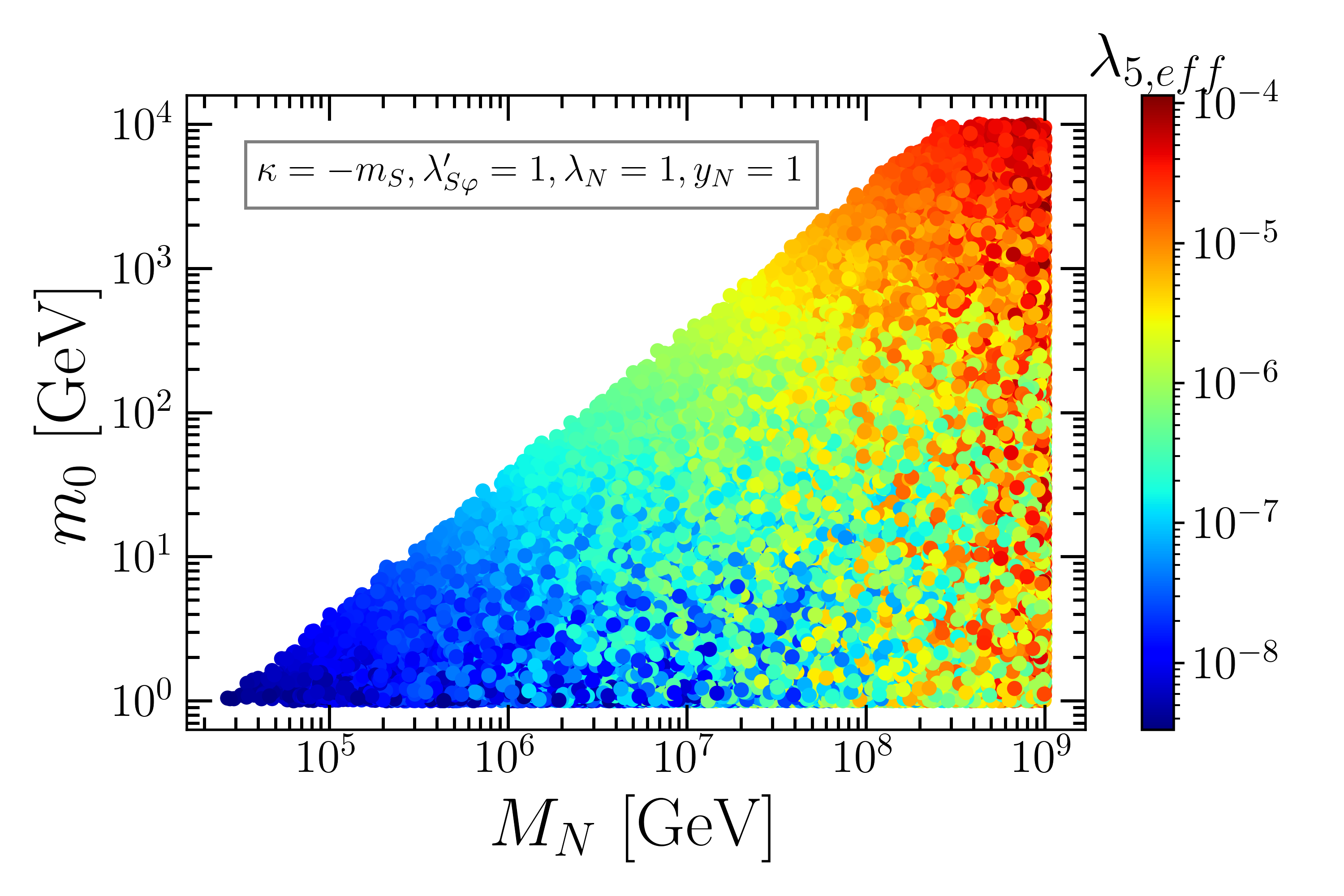}
\caption{Inert doublet scalar mass $m_0$ vs RH neutrino mass $M_N$ for varying $\lambda_{5,{\rm eff}}$. 
}
\label{fig:corr1}
\end{figure}

\begin{figure}[t]
\centering
\includegraphics[width=0.45\textwidth,clip]{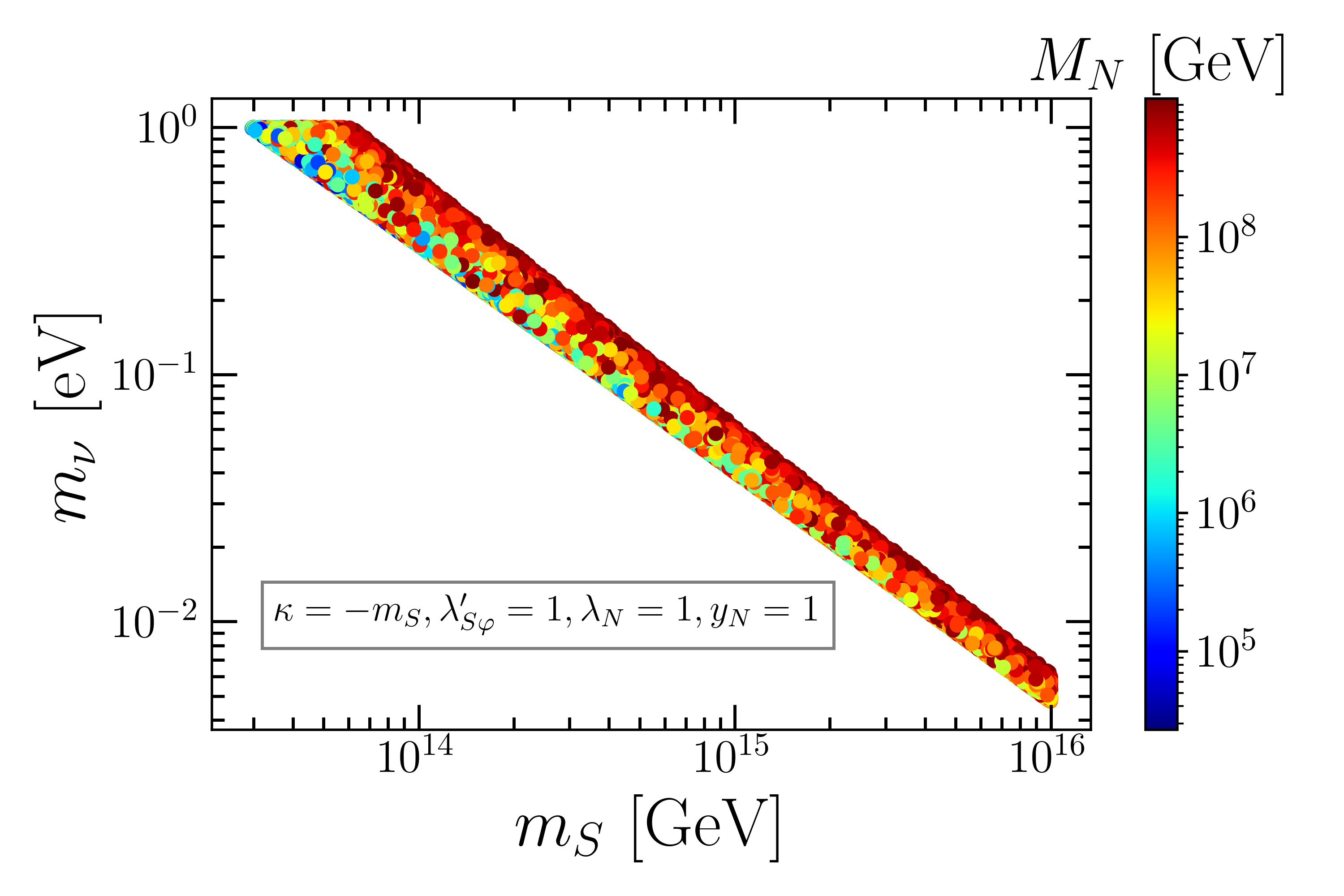}
\caption{Neutrino mass $m_\nu$ vs singlet scalar mass $m_S$ for varying $M_N$
}
\label{fig:corr2}
\end{figure}

In Fig.~\ref{fig:corr1}, we show the parameter space for inert doublet scalar mass $m_0$ vs RH neutrino mass $M_N\equiv M_{N,1}$ satisfying  $m_\nu=0.1\,{\rm eV}$. The color code indicates a scan over $\lambda_{5, {\rm eff}}$ between $10^{-8}$ and $10^{-4}$, showing that $\lambda_{5,{\rm eff}}\gtrsim 10^{-6}$ and $M_N\gtrsim 2\times 10^7\,{\rm GeV}$ are necessary for $m_0\gtrsim 100\,{\rm GeV}$.  
In Fig.~\ref{fig:corr2}, we decode the correlation between neutrino mass $m_\nu$ and singlet scalar mass $m_S$ for varying $M_N$ between $10^4\,{\rm GeV}$ and $10^9\,{\rm GeV}$. In both Figs.~\ref{fig:corr1} and \ref{fig:corr2}, we imposed $|\Delta|>100\,{\rm keV}$ and took into account the perturbativity, $|\lambda_{5,{\rm eff}}|<1$. Then, we need the singlet scalar mass $m_S$ to be of order $10^{14}-10^{16}\,{\rm GeV}$ for $m_\nu=0.01-0.1\,{\rm eV}$, which might hint at a connection of neutrino mass generation to the unification. 
If we choose smaller Yukawa couplings $y_N$, the diagonal boundary in Fig.~\ref{fig:corr1} is shifted to the left, allowing for smaller $M_N$ for a fixed $m_0$, and the strip in Fig.~\ref{fig:corr2} is shifted to the bottom, making $m_S$ smaller for a fixed $m_\nu$.

\section{ Dark matter direct detection}
When the lighter neutral scalar from the inert doublet is dark matter, we can avoid the strong bound from direct detection if the mass splitting between $A_0$ and $H_0$ is large enough, i.e. $|\Delta|\gtrsim 100\,{\rm keV}$. Then, the cross section for spin-independent scattering between dark matter and nucleus is dominated by the tree-level exchanges of Higgs-like scalars \cite{Barbieri:2006dq,LopezHonorez:2006gr}, as follows,
\bea
\sigma^{\rm SI}_{A_0}=\frac{\lambda_{L5}^2\mu^2_N(Z f_p+(A-Z)f_n)^2}{4\pi m^2_{A_0} A^2}\bigg(\frac{c^2_\alpha}{m^2_{h_1}}-\frac{s^2_\alpha}{m^2_{h_2}}\bigg)^2
\eea
where $\lambda_{L5}\equiv \lambda_L-\lambda_{5,\,{\rm eff}}$, $m_{h_{1,2}}$ are the masses for SM Higgs-like and dark Higgs-like scalars,  $Z, A-Z$ are the numbers of protons and neutrons in the detector nucleus, $\mu_N$ is the reduced mass for the DM-nucleus system, and $f_{p,n}$ are the nucleon form factors \cite{Hoferichter:2015dsa,Junnarkar:2013ac,Kim:2023pwf}. 

The DM relic density is determined by the DM annihilations into a pair of SM fermions, $WW, ZZ$ and $h_i h_j$ with $i,j=1,2$ \cite{Barbieri:2006dq,LopezHonorez:2006gr,Gong:2012ri}. If $H_0, H^\pm$ have comparable masses to $m_{A_0}$, we also need to include the co-annihilation channels associated with them as eq.~(\ref{Boltz}) in the appendix.  In particular, there is room to adjust $\lambda_L$ and dark Higgs-portal couplings, denoted by ${\cal L}_{\rm int}\supset -\frac{1}{2}\sum_{i=1,2}\lambda_{H_i\varphi}|H_i|^2\varphi^2$,  to explain the DM relic density for relatively heavy DM due to $A_0A_0\to h_i h_j$ while the direct detection bound is satisfied \cite{Belyaev:2016lok}. 
DM masses above about $100\,{\rm GeV}$ are consistent with the bounds from direct detection, as far as $|\lambda_{L5}(c^2_\alpha-s^2_\alpha m^2_{h_1}/m^2_{h_2})|\lesssim 10^{-2}$, apart from the resonance, and the bounds from the SM Higgs decays can be easily satisfied for $m_{A_0}, m_{h_2}\gtrsim m_{h_1}/2=62.5\,{\rm GeV}$. But, for a smaller value of $|\lambda_{L5}|$, as required near resonances, the one-loop corrections to the DM-nucleus cross section are relevant \cite{Abe:2015rja}. 

\begin{figure}[t]
\centering
\includegraphics[width=0.45\textwidth,clip]{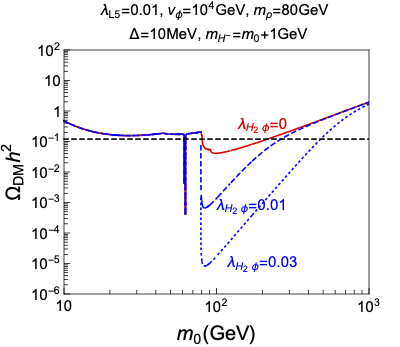}
\caption{DM relic density as a function of DM mass $m_0$, depending on the inert doublet portal coupling, $\lambda_{H_2\varphi}$. 
}
\label{fig:relic}
\end{figure}

In Fig.~\ref{fig:relic}, we depict the DM relic density as a function of DM mass for $\lambda_{H_2\varphi}=0, 0.01, 0.03$, in red solid, blue dashed and dotted lines, respectively. We chose a zero Higgs mixing, $\alpha=0$, the singlet parameters are $\lambda_{H_1\varphi}=0$,  $v_\varphi=10^4\,{\rm GeV}$ and $m_{h_2}=m_\rho=80\,{\rm GeV}$, and the parameters for the inert doublet are $\lambda_{L5}=0.01$, $\Delta=10\,{\rm MeV}$ and $m_{H^\pm}=m_0+1\,{\rm GeV}$. Thus, as $\lambda_{H_2\varphi}$ increases, the correct relic density can be explained for a heavier DM mass far beyond $100\,{\rm GeV}$. 
The results are insensitive to $\Delta$ as far as $\Delta\lesssim T_f$ where $T_f$ is the freeze-out temperature given by $T_f\simeq m_0/20$. 
Moreover, we remark that a nonzero $\lambda_{H_1\varphi}$ also contributes to the dark matter annihilation into $\varphi\varphi$ with the SM Higgs exchange, but its contribution is suppressed for a small $|\lambda_{L5}|$.

\section{ Thermal leptogenesis}
Baryogenesis can be realized in our model via thermal leptogenesis \cite{Fukugita:1986hr}, due to the decays of RH neutrinos with hierarchical masses. The RH neutrinos can be once in thermal equilibrium with the SM plasma in the early universe due to their Yukawa couplings to the inert doublet or the spurion scalar, which mixes with the SM Higgs. As the asymmetries generated from the decays of the heavier RH neutrinos, $N_{2,3}$, are negligible due to strong washout effects by scattering, we focus on the $N_1$ contribution to the lepton asymmetry.  The baryon-to-photon ratio converted from the final $B-L$ asymmetry is given by
\bea
\eta_B=-C\varepsilon_1 \kappa_1 
\eea
where $C\simeq 0.01$, $\varepsilon_1$ is the $CP$ asymmetry parameter, and $\kappa_1$ is the efficiency factor approximated \cite{Buchmuller:2004nz} to
\bea
\kappa_1(K_1)\simeq \frac{2}{z_B(K_1)K_1} \Big( 1-e^{-\frac{1}{2}z_B(K_1)K_1}\Big), 
\eea 
with $K_1=\Gamma_1/H(z_1=1)$ being the decay parameter and $z_B$ being the value of $z_1=M_{N,1}/T$ at the completion of the baryon asymmetry generation, given \cite{Buchmuller:2004nz} by
\bea
z_B(K)\simeq 1+\frac{1}{2} \ln\bigg(1+\frac{\pi K^2}{1024}\bigg[\ln \Big(\frac{3125\pi K^2}{1024}\Big)\bigg]^5\bigg).
\eea
Here, $\Gamma_1$ is the decay rate for $N_1$, given by $\Gamma_1\simeq \frac{M_{N,1}}{8\pi}(y^\dagger_N y_N)_{11}$ for $m_0\ll M_{N,1}$, and $H(z_1)$ is the Hubble parameter. In the Casas-Ibarra parametrization for the Yukawa matrix $y_N$ \cite{Casas:2001sr}, we can express $(y^\dagger_N y_N)_{ij}=\sqrt{\Lambda_i \Lambda_j} {\widetilde m}_{ij}$ in terms of the effective neutrino mass matrix ${\widetilde m}\equiv R D_{{\cal M}_\nu} R^\dagger$ with $R$ being arbitrary complex rotation matrix, $D_{{\cal M}_\nu}={\rm diag}(m_{\nu_1},m_{\nu_2},m_{\nu_3})$, and $\Lambda_i=\frac{4\pi^2\xi_i M_{N,i}}{\lambda_{5,{\rm eff}} v^2_H}$ \cite{Hugle:2018qbw}. Here, $\xi_i=\big(\frac{1}{8} \frac{M^2_{N,i}}{m^2_{H_0}-m^2_{A_0}}[L(m^2_{H_0})-L(m^2_{A_0})]\big)^{-1}$ with $L(m^2)=\frac{m^2}{m^2-M^2_{N,i}}\ln(m^2/M^2_{N,i})$ are order one parameters in most of the parameter space. Thus, we get the decay parameter as $K_1\simeq \frac{2\pi^2 \xi_1}{\lambda_{5\,{\rm eff}}}\sqrt{\frac{45}{64\pi^5 g_*}} \frac{M_P}{v^2_H}{\widetilde m}_{11}$ for $m_0\ll M_{N,1}$.

\begin{figure}[t]
\centering
\includegraphics[width=0.45\textwidth,clip]{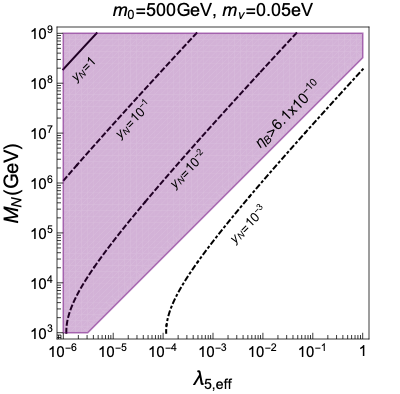}
\caption{Parameter space for RH neutrino mass $M_N$ vs $\lambda_{5,{\rm eff}}$, showing $\eta_B>6.1\times 10^{-10}$ in purple region. The contour lines show the parameter space for the heaviest neutrino mass equal to $m_\nu=0.05\,{\rm eV}$.
}
\label{fig:etaB}
\end{figure}

For the case with three RH neutrinos, the $CP$ asymmetry parameter is bounded \cite{Davidson:2002qv,Hugle:2018qbw} by
\bea
\varepsilon_1\lesssim \frac{3\pi}{4\lambda_{5\,{\rm eff}}v^2_H}\, \xi_3(m_h-m_l) M_{N,1}
\eea
with $m_h, m_l$ are the masses for the heaviest and lightest active neutrinos, respectively. 
After optimizing the efficiency factor $\kappa_1$ and the $CP$ asymmetry parameter $\varepsilon_1$, we obtain the relation, $\lambda_{5,{\rm eff}}\simeq 4\pi \xi_1 (m_l/10^{-3}\,{\rm eV})$ \cite{Hugle:2018qbw}, for which the baryon-to-photon ratio without  washout effects from $\Delta L$ processes is given by
\bea
\eta_B\simeq 3.8\times 10^{-10} \Big(\frac{M_{N,1}}{10^4\,{\rm GeV}}\Big) \bigg(\frac{10^{-4}}{\lambda_{5,{\rm eff}}}\bigg)\bigg(\frac{m_h}{0.1\,{\rm eV}}\bigg).
\eea

The efficiency factor $\kappa_1$ is subject to the corrections coming from the $\Delta L=2$ processes such as $l H_2\leftrightarrow {\bar l} {\bar H}_2$ and $ll\leftrightarrow {\bar H}_2 {\bar H}_2$.  Then, if the $\Delta L=2$ processes becomes important after the baryon asymmetry generation is finished, the total efficiency factor is given by $\kappa_{\rm tot}=\kappa_1\, {\rm exp}(-\int^\infty_{z_B} dz \Delta W)$ with $\Delta W=\Gamma_{\Delta L=2}/{H z_1}$ \cite{Buchmuller:2004nz,Hugle:2018qbw}. As $\Gamma_{\Delta L=2}$ is the scattering rate for the $\Delta L=2$ processes, we obtain $\int^\infty_{z_B} dz \Delta W=5.2 \times 10^{-13}\,{\rm GeV} \xi_3 M_{N,1} m_h M_P/(\sqrt{g_*}\lambda_{5,{\rm eff}} v^4_H)$ \cite{Hugle:2018qbw}, which is less than unity for $\lambda_{5,{\rm eff}}\gtrsim 1.7\times 10^{-6}$ for $M_{N,1}=10^4\,{\rm GeV}$ and $m_h=0.05\,{\rm eV}$. Then, in Fig.~\ref{fig:etaB}, using the analytic results, we show the parameter space for RH neutrino mass $M_N$ vs $\lambda_{5,{\rm eff}}$ where the maximum baryon-to-photon ratio for $m_h=0.05\,{\rm eV}$ is obtained as observed along the boundary of the purple region. The analysis in the region with $M_N\lesssim 10^5\,{\rm GeV}$ can be improved by numeral solutions to the Boltzmann equations \cite{Hugle:2018qbw}. For comparison, we also showed the contours for $m_h=0.05\,{\rm eV}$ with $m_0=500\,{\rm GeV}$, depending on the values of the Yukawa coupling, $y_N=10^{-3}-1$, in the same plot.

We remark that the asymmetry in the inert doublet or the DM asymmetry can reduce the lepton asymmetry, because $l H_2\to {\bar l} {\bar H}_2$ is favored over ${\bar l} {\bar H_2}\to l H_2$ \cite{Racker:2014yfa}. 
However, for a sizable $\lambda_{5,{\rm eff}}$, $H_2 {\bar H}_1\leftrightarrow {\bar H}_2 H_1$ process is fast, so it can make the DM asymmetry Boltzmann sufficiently suppressed. It turns out that $\Delta L=2$ processes are inefficient if $|\lambda_{5,{\rm eff}}|\gtrsim 10^{-5}$ for $M_{N,1}\lesssim 10^9\,{\rm GeV}$ \cite{Clarke:2015hta}.

\section{Conclusions}
We presented a unified solution with $Z_4$ discrete symmetry for small neutrino masses and stability of dark matter in the radiative seesaw scenarios. Once the $Z_4$ symmetry is broken to $Z_2$ by the VEV of the spurion singlet scalar $\varphi$ much below the mass of the singlet scalar $S$, a small effective lepton number violating interaction between two Higgs doublets is generated, being responsible for a small mass splitting of the neutral inert scalars at tree level and small neutrino masses at one-loop level. 

Although the $Z_2$ symmetry is sufficient for stability of dark matter, its extension to the $Z_4$ symmetry in our model allows for explaining the origin of small neutrino masses, even if inert doublet scalar and RH neutrinos are relatively light and the Yukawa couplings for neutrinos are sizable. This option gives rise to phenomenologically a novel window for new particle searches. 
Moreover, we showed that the spurion scalar responsible for the $Z_4$ symmetry breaking opens up a new parameter space for obtaining the correct relic density for inert doublet-like dark matter and a successful leptogenesis is achieved.

\textit{Acknowledgments}---

The work is supported in part by Basic Science Research Program through the National
Research Foundation of Korea (NRF) funded by the Ministry of Education, Science and
Technology (NRF-2022R1A2C2003567 and RS-2024-00341419 (JKK)).
This research was supported by the Chung-Ang University research grant in 2019.

\bibliographystyle{utphys}
\bibliography{ref2.bib}

\section{Appendix}
Here we provide the details on the full $Z_4$ invariant Lagrangian and discuss the case with a general mixing between the singlet scalar $S$ and the neutral components of $H_2$. We also discuss the process of determining the dark matter relic density in the limit of decoupling the singlet scalar $S$ while the excitation of a new singlet scalar $\varphi$ is kept. 

Nontrivial transformations of the fields under the $Z_4$ symmetry are given by $S\to i S, \varphi\to -\varphi, H_2\to i H_2$ and $N_R\to i N_R$, whereas the SM fields are neutral under $Z_4$. 

We first note that the $Z_4$ invariant scalar potential $V$ is composed of $V=V_1+V_2$, with 
\begin{widetext}
\begin{align}
V_1 &= m^2_1 \vert H_1 \vert^2 + m^2_2 \vert H_2 \vert^2 +\lambda_1 \vert H_1 \vert^4 +\lambda_2 \vert H_2 \vert^4 + \lambda_3 \vert H_1 \vert^2 \vert H_2 \vert^2 + \lambda_4 \left( H^\dagger_1 H_2 \right) \left( H^\dagger_2 H_1   \right)
\end{align}
\end{widetext}
and
\begin{widetext}
\bea
V_2&=& -\sqrt{2}\kappa S^\dagger H^\dagger_1 H_2 -\sqrt{2}\lambda'_{S\varphi}  \varphi S H^\dagger_1 H_2+\mu \, \varphi S^2+{\rm h.c.}\nonumber \\
&&+m^2_S \vert S\vert^2+\lambda_S \vert S\vert^4 +\frac{1}{2}m^2_\varphi\varphi^2 +\frac{1}{4}\lambda_\varphi  \varphi^4+\frac{1}{2} \lambda_{S\varphi} \vert S \vert^2 \varphi^2 + \sum_{i=1,2}\bigg[\lambda_{H_i S}\left(H^\dagger_i H_i \right) \vert S \vert^2  + \frac{1}{2}\lambda_{H_i \varphi}\left(H^\dagger_i H_i \right) \varphi^2\bigg]. \label{extrapot}
\eea
\end{widetext}

The relevant interaction Lagrangian for neutrino masses takes a more general form as compared to eq.~(\ref{Lint}), as follows,
\bea
{\cal L}_{\rm int}&=& -\sqrt{2}\kappa S^\dagger H^\dagger_1 H_2 -\sqrt{2}\lambda'_{S\varphi}  \varphi S H^\dagger_1 H_2+\mu \, \varphi S^2 \nonumber  \\
&&  -y_{N,ij} {\bar l}_i {\widetilde H}_2 N_{R,j} -\frac{1}{2}\lambda_{N,ij}\varphi \overline{N^c_{R,i}} N_{R,j} +{\rm h.c.} 
\eea
Then, after the $Z_4$ symmetry is spontaneously broken to $Z_2$ due to the VEV of $\varphi$, we obtain the effective interactions from eq.~(\ref{Lint}) as
\bea
{\cal L}_{{\rm eff}} &=& \sqrt{2}\kappa S^\dagger H^\dagger_1 H_2+\sqrt{2}\kappa' S H^\dagger_1 H_2+\frac{1}{2}{\hat m}^2_S S^2 \nonumber \\
&& -y_{N,ij} {\bar l}_i {\widetilde H}_2 N_{R,j} -\frac{1}{2}\lambda_{N,ij}\varphi \overline{N^c_{R,i}} N_{R,j} +{\rm h.c.}\label{Veff:Z2}
\eea
where $\kappa'=\lambda'_{S\varphi} v_\varphi$  and ${\hat m}^2_S=2 \mu v_\varphi$.  Here, the mass for the singlet $S$ is shifted by the contribution coming from $\lambda_{S\varphi} v^2_\varphi$, but we keep the notation for $m^2_S$ the same.
Moreover, integrating out the singlet scalar $S$ with ${\hat m}^2_S=2 \mu v_\varphi\ll \mu^2_S$ and $\kappa'\ll m_S$, from the combination of $\sqrt{2}\kappa S^\dagger H^\dagger_1 H_2, \sqrt{2}\kappa' S H^\dagger_1 H_2$ and $\frac{1}{2}{\hat m}^2_S S^2$ terms, we obtain the effective $\lambda_5$ term for two Higgs doublets, given by
\bea
\lambda_{5,{\rm eff}}=\frac{2{\hat m}^2_S (\kappa^2+\kappa^{\prime 2})}{m^4_S}-\frac{4\kappa\kappa'}{m^2_S}, \label{lambdaeff}
\eea
which generalizes the result in eq.~(\ref{lamb52}).
Thus, there are extra contributions due to the $\mu$ term for the interactions between $S$ and $\varphi$, but we can maintain a small value of the effective $\lambda_{5,{\rm eff}}$ coupling as far as ${\hat m}^2_S \ll m^2_S$, i.e. $v_\varphi\ll m_S$ for $\mu\sim m_S$.

\underline{\bf Minimization of the potential}

In unitarity gauge, two Higgs fields are expanded as
\begin{align}
H_1 = \frac{1}{\sqrt{2} } \begin{pmatrix} 0  \\ h+v_H 
\end{pmatrix},~~~~  H_2 &= \begin{pmatrix}
H^+ \\ \frac{1}{\sqrt{2}} \left(H_0 + i A_0 \right)
\end{pmatrix}.
\end{align}
On the other hand, the SM singlet scalar fields, $S$ and $\varphi$, are expanded as
\begin{align}
\varphi &= v_\varphi +\rho, ~~ S= \frac{1}{\sqrt{2}} \left( s + i a \right).
\end{align}
After the $Z_4$ symmetry is broken to $Z_2$, domain walls can be formed in the early universe, so we need an explicit breaking of the $Z_4$ symmetry with small soft mass terms or the $Z_4$ symmetry must be broken during inflation, in order to avoid domain wall problems.

We note that when the $Z_4$ symmetry is extended to $Z_{2n}$ with $n=3, 4,\cdots$, containing $Z_2$ as a subgroup or an approximate $U(1)$ global symmetry. In both cases, a similar mechanism as for $Z_4$ is working for ensuring lepton number conservation and stability of dark matter and we need to promote $\varphi$ to be a complex scalar field by $\varphi=v_\varphi +\rho+i\eta$ and change $\varphi$ interactions in the previous discussion appropriately. Then, in the case of an approximate $U(1)$ global symmetry, there can be phenomenologically interesting consequences due to a pseudo-Goldstone boson $\eta$. 

Then, to determine the minimum of the potential, we consider the potential only for the CP-even scalars, as follows,
\begin{widetext}
\bea
V&=&\frac{1}{2} m^2_1 h^2 + \frac{1}{2} m^2_2 (H_0)^2 + \frac{1}{4} \lambda_1 h^4 + \frac{1}{4} \lambda_2 (H_0)^4 +\frac{1}{4} (\lambda_3+\lambda_4) h^2 (H_0)^2 \nonumber \\
&&+\kappa s H_0 h + \frac{1}{2} \lambda'_{S\varphi} \rho h s H_0 +\mu \rho s^2 + \frac{1}{2} m^2_S s^2 +\frac{1}{4} \lambda_S s^4 + \frac{1}{2} m^2_\varphi \rho^2 + \frac{1}{4} \lambda_\varphi \rho^4 \nonumber \\
&&+ \frac{1}{4} (\lambda_{H_1 S}h^2+ \lambda_{H_2 S} (H_0)^2) s^2 + \frac{1}{4} (\lambda_{H_1\varphi} h^2+ \lambda_{H_2\varphi} (H_0)^2 )\rho^2+\frac{1}{4}\lambda_{S\varphi} s^2 \rho^2. 
\eea
\end{widetext}

Here, we can determine the VEVs by imposing the first derivatives of the CP-even scalar fields to vanish, i.e. $\frac{\partial V}{\partial \chi_i}$=0, for $\chi_i=h, \rho, H_0, s$, with
\begin{align}
\frac{\partial V}{\partial h} &= m^2_1h +\lambda_1 h^3  + \frac{1}{2}(\lambda_3+\lambda_4)h (H_0)^2+\kappa sH_0  \nonumber\\
&\quad+\frac{1}{2}\lambda'_{S\varphi}\rho s H_0+ \frac{1}{2} \lambda_{H_1 S} h s^2 + \frac{1}{2} \lambda_{H_1 \varphi}h \rho^2,\\
\frac{\partial V}{\partial \rho} &= m^2_\varphi \rho +\lambda_\varphi \rho^3 +\frac{1}{2}\lambda'_{S\varphi} h s H_0 + \mu s^2  \nonumber\\
&\quad+ \frac{1}{2}\lambda_{H_1 \varphi}h^2\rho + \frac{1}{2}\lambda_{H_2 \varphi}(H_0)^2\rho  +\frac{1}{2} \lambda_{S\varphi} s^2\rho , \\
\frac{\partial V}{\partial H_0} &= m^2_2 H_0 +\lambda_2 (H_0)^3 +\frac{1}{2}(\lambda_3+\lambda_4)h^2 H_0 +\kappa s h \nonumber \\
&\quad + \frac{1}{2}\lambda'_{S\varphi} \rho h s +\frac{1}{2}\lambda_{H_2S} H_0 s^2 + \frac{1}{2}\lambda_{H_2\varphi} H_0 \rho^2,   \\  
\frac{\partial V}{\partial s} &= m^2_S s+\lambda_S s^3 +\kappa H_0 h + \frac{1}{2}\lambda'_{S\varphi} \rho h H_0 +2 \mu \rho s\nonumber \\
&\quad + \frac{1}{2} \lambda_{H_1S} h^2 s + \frac{1}{2} \lambda_{H_2S} (H_0)^2 s + \frac{1}{2}\lambda_{S\varphi}s\rho^2.
\end{align}

We now take the VEVs of $h$ and $\rho$ fields to $v_H$ and $v_\varphi$, respectively, whereas setting the VEVs of the other fields  to zero.
Then, the minimization conditions for the potential become
\begin{align}
m^2_1 &= -\lambda_1 v_H^2 - \frac{1}{2}\lambda_{H_1 \varphi} v^2_\varphi , \\
m^2_\varphi &=  - \lambda_{\varphi} v^2_\varphi - \frac{1}{2}\lambda_{H_1 \varphi} v^2_H.
\end{align}
Therefore, we can determine the VEVs as
\bea
v_H &=& 2\,\frac{\lambda_{H_1 \varphi} m^2_\varphi-2\lambda_\varphi m^2_1 }{4\lambda_1\lambda_\varphi-\lambda_{H_1 \varphi}^2}, \label{vH} \\
v_\varphi &=&2\,\frac{\lambda_{H_1 \varphi} m^2_1-2\lambda_1 m^2_\varphi }{4\lambda_1\lambda_\varphi-\lambda_{H_1 \varphi}^2}. \label{vphi}
\eea

By using the results in eqs.~(\ref{vH}) and (\ref{vphi}), the diagonal elements of the squared mass matrix for $(h,\rho)$, ${\cal M}^2_{h,\rho}$, become $2\lambda_1 v^2_H$ and $2\lambda_\varphi v^2_\varphi$, respectively, so  ${\rm det}({\cal M}^2_{h,\rho})=(4\lambda_1\lambda_\varphi-\lambda_{H_1 \varphi}^2) v^2_H v^2_\varphi>0$.
Thus, together with eqs.~(\ref{vH}) and (\ref{vphi}), we find the conditions for a local minimum as
\bea
&&4\lambda_1\lambda_\varphi-\lambda_{H_1 \varphi}^2>0, \\
&& \lambda_{H_1 \varphi} m^2_\varphi-2\lambda_\varphi m^2_1>0, \\
&&\lambda_{H_1 \varphi} m^2_1-2\lambda_\varphi m^2_\varphi>0. 
\eea

\underline{\bf Higgs mixing}

The SM Higgs mixes with the neutral component of singlet scalar $\varphi$ from the mass matrix, given by
\begin{align}
{\cal M}^2_{h, \rho}=\begin{pmatrix} m^2_{h} &  \lambda_{H_1\varphi} v_H v_\varphi \\  \lambda_{H_1\varphi}  v_H v_\varphi& m^2_\rho   \end{pmatrix}
\end{align}
where $m^2_h =2\lambda_1 v^2_H$ and $m^2_\rho=2\lambda_\varphi v^2_\varphi$.
Then, we can diagonalize the mass matrix by introducing the mixing matrices, as follows,
\begin{align}
	\begin{pmatrix}  h \\ \rho \end{pmatrix} =  \begin{pmatrix}  \cos \alpha & -\sin \alpha \\ \sin \alpha & \cos \alpha \end{pmatrix} \begin{pmatrix} h_1 \\ h_2 \end{pmatrix}
\end{align}
with
\begin{align}
\tan 2\alpha &= {2v_H v_\varphi \lambda_{H_1\varphi} \over m^2_{h}- m^2_\rho }.
\end{align} 
Then, the mass eigenvalues for neutral scalars are given by
{\small
\bea
m^2_{h_{1, 2}} &=&\frac{1}{2}\bigg[m^2_{h}+m^2_\rho \pm (m^2_{h}-m^2_\rho) \sqrt{1+\frac{4 v_H^2 v_\varphi^2 \lambda_{H_1\varphi}^2 }{(m^2_{h}-m^2_\rho)^2 }} \bigg].
\eea
}

\underline{\bf Dark matter mixing}

Without mixing mass terms, we first list the scalar masses for the inert Higgs doublet and the real scalars, $s, a$, of the singlet $S$ as
\begin{align}
m^2_{H^\pm} &= m^2_2 +\frac{1}{2 }\lambda_3 v^2_H+ \frac{1}{2}\lambda_{H_2\varphi} v^2_\varphi, \label{HCmass} \\
m^2_{H_0} &=   m^2_2 + \frac{1}{2} \left( \lambda_3+\lambda_4 \right)v^2_H +\frac{1}{2} \lambda_{H_2\varphi} v^2_\varphi , \\
m^2_{A_0}&= m^2_2 + \frac{1}{2} \left( \lambda_3+\lambda_4 \right)v^2_H +\frac{1}{2} \lambda_{H_2\varphi} v^2_\varphi , \\
m^2_s&=m^2_S +\frac{1}{2}\lambda_{H_1S} v^2_H +{\hat m}^2_S +\frac{1}{2} \lambda_{S\varphi} v^2_\varphi, \\
m^2_a&=m^2_s-2{\hat m}^2_S.
\end{align}
In the presence of the unbroken $Z_4$ symmetry, the masses of $H_0$ and $A_0$ fields are the same, that is, $m^2_{H_0}=m^2_{A_0}$.
We note that that a negative $\lambda_4$ is taken for the extra neutral Higgs scalars to be lighter than the charged Higgs.

In our model, however, there are mass mixings between the neutral scalars of the inert doublet and the singlet scalars containing in $S$, leading to split masses between neutral scalar fields.
First, the mass matrix for CP-even scalars, $(H_0, s)$, is given by
\begin{align}
{\cal M}^2_{H_0, s}=\begin{pmatrix} m^2_{H_0} & (\kappa+\kappa') v_H \\ (\kappa+\kappa') v_H  & m^2_s   \end{pmatrix}.
\end{align}
We also obtain the mass matrix  for the CP-odd scalars, $(A_0, a)$,  as
\begin{align}
{\cal M}^2_{A_0, a}=\begin{pmatrix} m^2_{H_0} & (\kappa-\kappa') v_H \\ (\kappa-\kappa') v_H  & m^2_a  \end{pmatrix}.
\end{align}
Then, we can diagonalize the mass matrices by introducing the mixing matrices, as follows,
\bea
\begin{pmatrix}  H_0 \\ s \end{pmatrix} =  \begin{pmatrix}  \cos\theta_s  & -\sin\theta_s \\ \sin\theta_s & \cos\theta_s \end{pmatrix} \begin{pmatrix} H_1 \\ H_2 \end{pmatrix}, 
\eea
\bea
\begin{pmatrix}  A_0 \\ a \end{pmatrix} =  \begin{pmatrix}  \cos\theta_a  & -\sin\theta_a\\ \sin\theta_a & \cos\theta_a \end{pmatrix} \begin{pmatrix} A_1 \\ A_2 \end{pmatrix},
\eea
with
\begin{align}
\tan 2\theta_s &=  {2(\kappa+\kappa') v_H \over m^2_{H_0}-m^2_s }, \\
\tan 2\theta_a &= {2(\kappa-\kappa') v_H \over m^2_{H_0}- m^2_a }.
\end{align} 
Then, the mass eigenvalues for neutral scalars are given by
\begin{widetext}
\bea
m^2_{H_{1,2}} &=&\frac{1}{2}\bigg[m^2_{H_0}+m^2_s\pm (m^2_{H_0}-m^2_s) \sqrt{1+\frac{4(\kappa+\kappa')^2 v^2_H}{(m^2_{H_0}-m^2_s)^2 }} \bigg],  \label{Hmasses} \\
m^2_{A_{1,2}} &=&\frac{1}{2}\bigg[m^2_{H_0}+m^2_a\pm (m^2_{H_0}-m^2_a) \sqrt{1+\frac{4(\kappa-\kappa')^2 v^2_H}{(m^2_{H_0}-m^2_a)^2 }} \bigg].
\label{Amasses}
\eea
\end{widetext}
Therefore, the lightest neutral scalar among $H_{1,2}$ and $A_{1,2}$ becomes a candidate for dark matter. In this case, even if $m^2_{H_0}=m^2_{A_0}$ due to the $Z_4$ symmetry, there arises a mass splitting between the CP-even and CP-odd scalars due to $m^2_s\neq m^2_a$ and $\kappa'\neq 0$, which stem from the $Z_4$ breaking mass terms for the singlet scalar. So, when the mass splitting is larger than about $ 100$ keV, we can avoid the strong direct detection bound on the inert doublet-like DM in the case of the inelastic DM-nucleon scattering mediated by $Z$ boson.

\underline{\bf General formulas for neutrino masses}

In the case with the non-decoupled singlet scalar $S$, we present the general results for one-loop neutrino masses in our model. 
From eq.~(\ref{Leff}), the relevant interactions for neutrino masses are
\begin{widetext}
\bea
{\cal L}_{\rm \nu,{\rm eff}}
=- \frac{1}{\sqrt{2}}y_{N,ij} {\bar\nu}_i \Big(\cos\theta_s H_1-\sin\theta_s H_2-i\cos\theta_a A_1+i\sin\theta_a A_2\Big) N_{R,j} -\frac{1}{2}M_{N,i} \overline{N^c_{R,i}} N_{R,i} +{\rm h.c.} 
\eea
\end{widetext}
As a result, we obtain the neutrino masses from one-loop diagrams as
\begin{widetext} 
\bea 
({\cal M}_\nu)_{ij}=\frac{1}{16\pi^2}\sum_\alpha\sum_{k} y_{N,ik} y_{N,jk} M_{N,k} F_\alpha \bigg[\frac{m^2_\alpha}{m^2_\alpha-M^2_{N,k}}\ln \frac{m^2_\alpha}{M^2_{N,k}} \bigg]  \label{genumass}
\eea
\end{widetext}
with $F_\alpha\equiv(\cos^2\theta_s,\sin^2\theta_s,-\cos^2\theta_a, -\sin^2\theta_a )$ and $m_\alpha\equiv(m_{H_1},m_{H_2},m_{A_1},m_{A_2})$.
Thus, in the limit of decoupling the singlet scalar $S$ with $m_S\gg \kappa', {\hat m}_S$,  the mixing angles for dark scalars become $|\theta_s|, |\theta_a|\ll 1$ and the mass splittings between dark scalars become small. In this case, from eqs.~(\ref{Hmasses}) and (\ref{Amasses}), we obtain the mass splitting between the neutral inert doublet-like scalars as
\bea
&&m^2_{H_1}-m^2_{A_1} \simeq  \frac{(\kappa+\kappa')^2 v^2_H}{m^2_{H_0}-m^2_s}-\frac{(\kappa-\kappa')^2 v^2_H}{m^2_{H_0}-m^2_a} 
\nonumber \\
&&\quad=\frac{2(\kappa^2+\kappa^{\prime 2}){\hat m}^2_S v^2_H + 4\kappa\kappa' (m^2_{H_0}-m^2_{s,0})v^2_H}{(m^2_{H_0}-m^2_s)(m^2_{H_0}-m^2_a)},
\eea
with $m^2_{s,0}\equiv m^2_s-{\hat m}^2_S$. As $m^2_s, m^2_a \simeq m^2_S\gg m^2_{H_0}$, the above result becomes further approximated to
\bea
m^2_{H_1}-m^2_{A_1} \simeq \frac{2(\kappa^2+\kappa^{\prime 2}){\hat m}^2_S v^2_H}{m^4_S} -\frac{4\kappa\kappa' v^2_H}{m^2_S}, \label{eq:massdeiff}
\eea
which is equal to $\lambda_{5,{\rm eff}} \, v^2_H$, being consistent with the effective $\lambda_{5,{\rm eff}}$ term in eq.~(\ref{lambdaeff}), obtained after the singlet scalar is integrated out.
Therefore, in the decoupling limit of the singlet scalar $S$, the neutral inert doublet-like scalars contribute dominantly to the one-loop neutrino masses  as
\begin{widetext}
\bea
({\cal M}_\nu)_{ij} \simeq \frac{\lambda_{5,{\rm eff}}v^2_H}{16\pi^2}\sum_k \frac{y_{N,ik}y_{N,jk}M_{N,k}}{m^2_0-M^2_{N,k}}\bigg[1-\frac{M^2_{N,k}}{m^2_0-M^2_{N,k}}\ln\frac{m^2_0}{M^2_{N,k}}\bigg],
\eea
\end{widetext}
where $m^2_0=(m^2_{H_1}+m^2_{A_1})/2\simeq m^2_{H_0}$ and  $\lambda_{5,{\rm eff}}$ is given by eq.~(\ref{lambdaeff}). We remark that, without accidental cancellations in $\lambda_{5,{\rm eff}}$, the decoupling limit for the singlet scalar $S$ is desirable for small neutrino masses.
On the other hand, for sizable dark mixing angles for which the singlet scalar $S$ is not decoupled, the mass splittings between dark scalars become large, so small neutrino masses require small Yukawa couplings or large masses for the RH neutrinos.

\underline{\bf Dark matter density}

When $A_0$ is the lightest neutral $Z_2$-odd particle, we need to determine the relic density by solving the Boltzmann equations for the number density of $A_0$. But, when the other scalars in the inert doublet have comparable masses to the one for $A_0$, it is important to include the co-annihilation processes for them too.

Suppose that the singlet $Z_2$-odd scalar $S$ is decoupled but the excitation of the singlet scalar $\varphi$ is kept in the low energy. 
Ignoring the Higgs mixing, we obtain the Boltzmann equation governing the number density for $A_0$,  $n_{A_0}$, as
\bea
&&\frac{d n_{A_0}}{dt} + 3H n_{A_0} = -\langle \sigma v\rangle_{2,{\rm eff}} \Big(n^2_{A_0}-(n^{\rm eq}_{A_0})^2\Big) \nonumber \\
&&-\langle \sigma v\rangle_{A_0 H_0\to f{\bar f}} \Big(n_{A_0}n_{H_0}-n^{\rm eq}_{A_0} n^{\rm eq}_{H_0}\Big) \nonumber \\
&&-\langle \sigma v\rangle_{A_0 H^-\to f{\bar f}'}  \Big(n_{A_0}n_{H^-}-n^{\rm eq}_{A_0} n^{\rm eq}_{H^-}\Big)  \nonumber \\
&&-\langle \sigma v\rangle_{A_0 H^+\to f'{\bar f}}  \Big(n_{A_0}n_{H^+}-n^{\rm eq}_{A_0} n^{\rm eq}_{H^+}\Big) 
\label{Boltz}
\eea
where
\bea
\langle \sigma v\rangle_{2,{\rm eff}}&=&  \langle \sigma v\rangle_{A_0 A_0\to hh} + \langle \sigma v\rangle_{A_0 A_0\to h\rho} \nonumber \\
&&+ \langle \sigma v\rangle_{A_0 A_0\to \rho\rho}+ \langle \sigma v\rangle_{A_0 A_0\to WW}\nonumber \\
&&+\langle \sigma v\rangle_{A_0 A_0\to ZZ}+\langle \sigma v\rangle_{A_0 A_0\to f{\bar f}}.
\eea
Here, we note that there is a factor of $1/2$ included in each thermal averaged cross section for two identical particles, which cancels a factor of $2$ in each annihilation rate for two identical particles.
There are similar Boltzmann equations for the heavier scalars, $H_0, H^\pm$, with the corresponding annihilation cross sections.

In the presence of particles with similar masses, participating in the co-annihilation processes, the heavier scalars, $H_0, H^\pm$, that survive annihilation, eventually decay into $A_0$. Thus, we consider the Boltzmann equation for the total number density of $A_0, H_0, H^\pm$, participating in the (co-)annihilation processes, i.e. $n_{\rm DM}=n_{A_0}+n_{H_0}+n_{H^\pm}$, in the following form,
\bea
\frac{dn_{\rm DM}}{dt} +3 Hn_{\rm DM} = -\sum_{a,b} \langle\sigma v\rangle_{ab} (n_a n_b-n^{\rm eq}_a n^{\rm eq}_b),
\eea
with $a,b=A_0, H_0, H^\pm$ being annihilating particles.
Making an approximation for each component in the total number density by $n_a/n_{\rm DM}\simeq n^{\rm eq}_a/n^{\rm eq}_{\rm DM}\equiv w_a/w$, with
\bea
w_a=g_a (1+\Delta_a)^{3/2} e^{-x\Delta_a}, \,\, \Delta_a\equiv \frac{m_a-m_{A_0}}{m_{A_0}},
\eea
and $w=\sum_a w_a$,  we can rewrite the above Boltzmann equation as
\bea
\frac{dn_{\rm DM}}{dt} +3 Hn_{\rm DM} = -\langle\sigma v\rangle_{\rm eff} \Big(n^2_{\rm DM} -(n^{\rm eq}_{\rm DM})^2\Big)
\eea
where the effective annihilation cross section is given by
\bea
(\sigma v)_{\rm eff}=\frac{1}{w^2}\sum_{a,b} (\sigma v)_{ab} w_a w_b.
\eea
We note that $g_a=1$ for $a=A_0, H_0, H^+, H^-$. 

Then, the relic abundance for dark matter is determined by
\bea
\Omega_{\rm DM} h^2=0.2744\bigg(\frac{Y_{\rm DM}}{10^{-11}}\bigg)\bigg(\frac{m_{A_0}}{100\,{\rm GeV}}\bigg)
\eea
where $Y_{\rm DM}=n_{\rm DM}/s$ is the present abundance for dark matter, given by
\bea
Y_{\rm DM}=\sqrt{\frac{45 g_*}{8\pi^2 g^2_{*s}}} \,\frac{x_f}{m_{A_0}M_P (I_s + 3 I_p/x_f) },
\eea
where we made a velocity expansion by $(\sigma v)_{ij}=a_{ij}+b_{ij} v^2$ and $(\sigma v)_{\rm eff}=a_{\rm eff}+b_{\rm eff} v^2$, and
\bea
I_s=x_f \int^\infty_{x_f} x^{-2} a_{\rm eff} dx, \,\,  I_p= 2x^2_f \int^\infty_{x_f} x^{-3} b_{\rm eff} dx.
\eea
Here,  we note that 
{\small
\bea
x_f=\ln y_f-\frac{1}{2} \ln \ln y_f, \, y_f= \sqrt{\frac{45}{4\pi^5 g_*}} m_{A_0} M_P \langle\sigma v\rangle_{\rm eff},
\eea 
}
and $M_P$ is the reduced Planck scale, given by $M_P=2.435\times 10^{18}\,{\rm GeV}$,  $g_*=3.36$ and $g_{*s}=3.91$.

If $ \Delta_a\gtrsim 1/x_f$ for $a=H_0, H^\pm$, we can ignore the (co-)annihilation processes for the heavier scalars, $H_0, H^\pm$, because of the Boltzmann suppression of the number densities for the heavier scalars. In this case, $a_{\rm eff}\simeq a_{A_0 A_0}$ and $b_{\rm eff}\simeq b_{A_0A_0}$, so $I_s=I_p=1$.

If $ \Delta_{H^\pm}\gtrsim 1/x_f$ but $\Delta_{H_0}\lesssim 1/x_f$, we need to keep the heavier scalar $H_0$ in the total number density for dark matter, so we get 
\bea
a_{\rm eff}\simeq \frac{1}{4}(a_{A_0 A_0} + 2 a_{A_0 H_0}+a_{H_0 H_0})
\eea
and  $b_{\rm eff}$ with $a_{ij}\to b_{ij}$. In this case, we obtain $I_s=\frac{1}{4}(a_{A_0 A_0} + 2 a_{A_0 H_0}+a_{H_0 H_0})/a_{A_0A_0}$ and $I_p=\frac{1}{4}(b_{A_0 A_0} + 2 b_{A_0 H_0}+b_{H_0 H_0})/b_{A_0A_0}$.

If $ \Delta_{H_0}\lesssim \Delta_{H^\pm}\lesssim 1/x_f$, we need to include the (co-)annihilation processes for the heavier scalars in determining the total number density for dark matter, so we get
\bea
&&a_{\rm eff}\simeq \frac{1}{4(1+w_{H^-})^2}\Big(a_{A_0 A_0} + 2a_{A_0 H_0}+a_{H_0 H_0} \nonumber \\
&&+4w_{H^-}a_{A_0 H^-}+4w_{H^-}a_{H_0 H^-}+2w^2_{H^-}a_{H^+ H^-}\Big)
\eea
and   $b_{\rm eff}$ with $a_{ij}\to b_{ij}$.

\end{document}